\documentclass{article}
\usepackage{spconf}
\usepackage{hyperref}
\usepackage{amsmath}
\usepackage{amsfonts}
\usepackage{amssymb}
\usepackage{amscd}
\usepackage[dvips]{graphicx}
\usepackage{tikz}

\usepackage{amsthm}
\usepackage{newlfont}
\usepackage{color}
\usepackage{setspace}
\usepackage{flushend}
\newcommand{\beq}{\begin{equation}}
\newcommand{\eeq}{\end{equation}}
\newcommand{\beqa}{\begin{eqnarray}}
\newcommand{\eeqa}{\end{eqnarray}}

\usepackage{cite}

\title{Human-Machine Inference Networks for Smart Decision Making: Opportunities and Challenges}
\name{Aditya Vempaty$^\dagger$  \qquad Bhavya Kailkhura$^\ddagger$ \qquad Pramod K. Varshney$^\ast$
 \address{$^\dagger$ IBM T. J. Watson Research Center, Yorktown Heights, NY\\
 $^\ddagger$ Lawrence Livermore National Laboratories, Livermore, CA\\
 $^\ast$ Department of Electrical Engineering and Computer Science, Syracuse University, Syracuse, NY  }
 \thanks{This work was performed under the auspices of the U.S. Department of Energy by Lawrence Livermore National Laboratory under Contract DE-AC52-07NA27344 (LLNL-CONF-742883) and supported in part by the AFOSR DDDAS program under grant FA9550-17-1-0313. }
 }

\begin{document}

\maketitle
\begin{abstract}
The emerging paradigm of Human-Machine Inference Networks (HuMaINs) combines complementary cognitive strengths of humans and machines in an intelligent manner to tackle various inference tasks and achieves higher performance than either humans or machines by themselves. While inference performance optimization techniques for human-only or sensor-only networks are quite mature, HuMaINs require novel signal processing and machine learning solutions. In this paper, we present an overview of the HuMaINs architecture with a focus on three main issues that include architecture design, inference algorithms including security/privacy challenges, and application areas/use cases.

\end{abstract}


\section{Introduction}
In traditional economics, cognitive psychology, and artificial intelligence (AI) literature, the problem-solving or inference process is described in terms of searching a problem space, which consists of various states of the problem, starting with the initial state and ending at the goal state which one would like to reach \cite{Anderson2010}. Each path from the initial state represents a possible strategy which can be used. These paths could either lead to the desired goal state or to other non-goal states. The paths from the initial state that lead to the goal state are called the solution paths. There could be multiple such paths between the initial and the goal state which are all solutions to the problem. In other words, there are multiple ways to solve a given problem. The problem-solving process is to identify the optimal (under a given constraint) solution path among the multiple solution paths emanating from the initial state and reaching the goal state. 

The first step for such a search is to determine the set of available strategies, i.e., the strategy space. The second step is to evaluate the strategies to determine the best strategy as the solution. In traditional economic theory, a rational decision maker is assumed to have the knowledge of the set of possible alternatives\footnote{The terms strategy and alternative are used interchangeably.}, has the capability to evaluate the consequences of each alternative, and has a utility function which he/she tries to maximize to determine the optimal strategy \cite{MarchS1958}. However, it is widely accepted that humans are not rational but are bounded rational agents. Under the bounded rationality framework \cite{MarchS1958,Simon1982}, decision makers are cognitively limited and have limited time, limited information, and limited resources. The set of alternatives is not completely known \textit{a priori} nor are the decision makers perfectly aware of the consequences of choosing a particular alternative. Therefore, the decision maker might not always determine the best strategy for solving the problem. 

On the other hand, machines\footnote{The terms sensor and machine are used interchangeably.} are \emph{rational} in the sense that they have stronger/larger memory for storing alternatives and have the computational capability to more accurately evaluate the consequences of a particular alternative. Therefore, a machine can aid a human in fast and accurate problem-solving. This leads us to a framework for human-machine collaboration for problem-solving. In this paper, we discuss this collaboration framework for inference by defining the Human-Machine Inference Networks (HuMaINs) and discuss the research challenges associated with developing such a framework. The three basic threads of research in this area are defined. 

\section{HuMaIN framework}
\label{sec:humain}
\begin{figure}[htbp]
\includegraphics[width=0.475\textwidth]{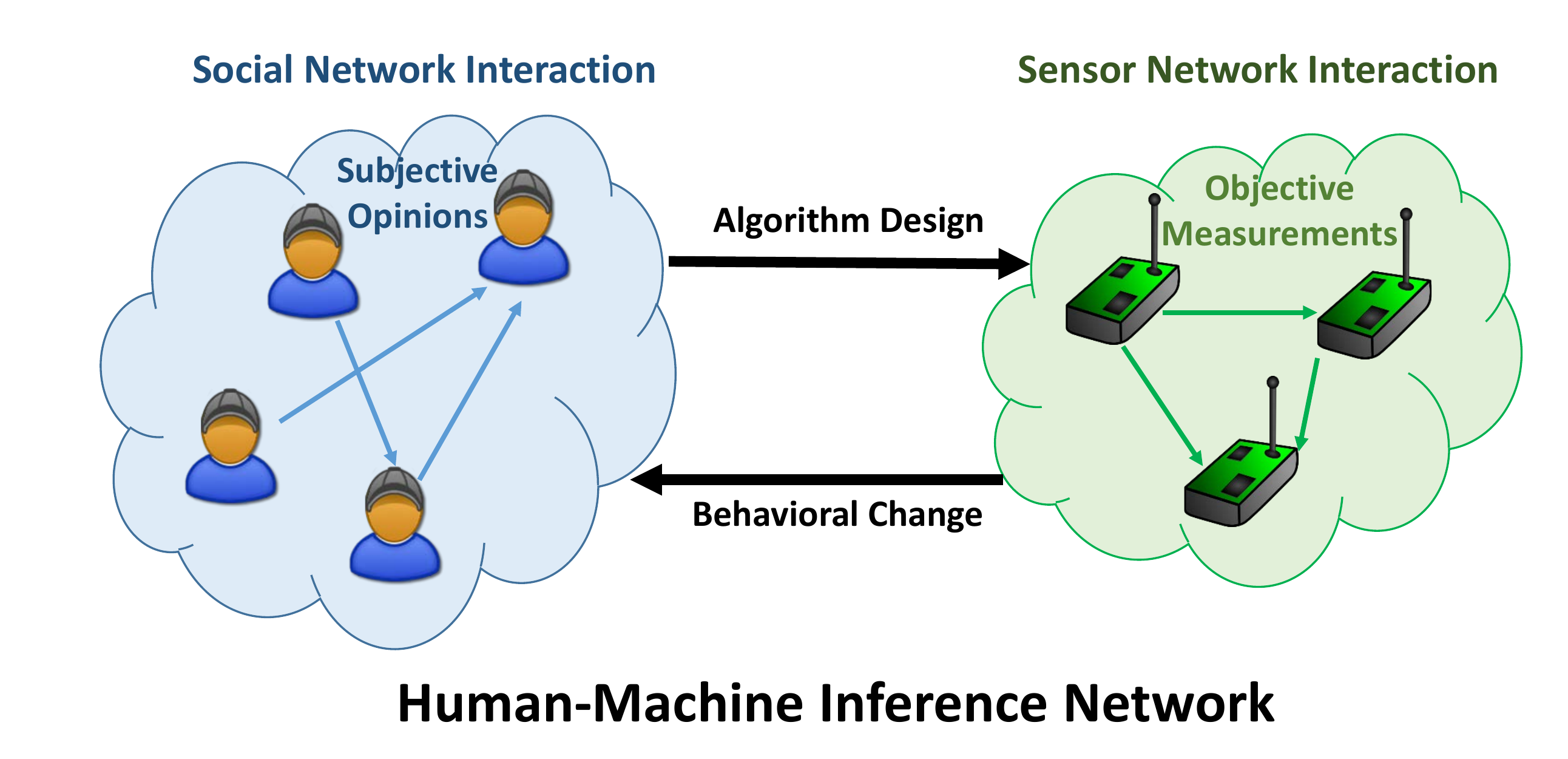}
\caption{Notional HuMaINs architecture}
\label{fig:framework}
\end{figure}

Fig.~\ref{fig:framework} presents a typical Human-Machine Inference Network (HuMaIN). A typical HuMaIN consists of a social network where humans exchange subjective opinions among themselves, and a machine network where machines exchange objective measurements amongst them. Moreover, due to the interaction between social and machine networks, the behavioral characteristics of humans determine algorithms adopted by machines and these algorithms in turn affect the behavior of humans. Therefore, an intelligent collaboration of humans and machines can deliver improved results, by exploiting the strengths of humans and machines.

There are three major directions of research that fall under the HuMaIN paradigm: $1)$ architecture, $2)$ algorithms, and $3)$ applications.

\section{Architecture}
Several control architectures involve the interaction of an autonomous system with one or more human agents. Examples of such architecture include fly-by-wire aircraft control systems (interacting with a pilot), automobiles with driver assistance systems (interacting with a driver), and medical devices
(interacting with a doctor, nurse, or patient)~\cite{LiSSS2014}. 
The success of such architecture depends not only on the autonomous system, but also on the actions of the human agents. 
The goal is to develop a human decision-making framework that quantifies the human representation in the decision-making task under uncertainty, and also develop an estimator for the model parameters.
The framework should also provide a common ontology for humans and machines to share relevant information about the task. By estimating the parameters, a machine can access this representation and potentially improve its performance. In control systems terminology, the model and associated estimator should form a plant-observer pair for human decision making that can be used for system design~\cite{Reverdy2014}.
Incorporating these ideas into the feedback control framework will require new results and theory to provide performance guarantees. 

We classify the architectures into three categories: $1)$ architectures where humans directly control the autonomous system, $2)$ architectures where the autonomous system monitors humans and takes actions if required, and $3)$ a combination of $1$ and $2$. 
In order to achieve the goal of HuMaINs, it is critical to build an architecture that lends itself to a blend of human and machine decision making. In \cite{PretloveS2007}, it is stated that to create a state-of-the-art operator environment for modern automation systems, continued technology development is needed in three major areas: decision support tools; ergonomics and visualization technologies; and ease-of-use of complex systems. Research focusing on building such systems falls under the research paradigm of Human-in-the-loop cyber-physical system (HiLCPS) \cite{SchirnerECP2013}. As Schirner et al. \cite{SchirnerECP2013} state, designing and implementing a HiLCPS poses challenges that requires multi-disciplinary research to solve these challenges. Research in the areas of control systems, human-computer interface (HCI), and systems design, together will drive the design of a HuMaIN architecture.

\section{Algorithms}
The key research area for HuMaINs is the development of new algorithms that deal with the human-behavioral data. This falls under the paradigm of an emerging research area called \emph{behavioral signal processing} \cite{NarayananG2013}. Behavioral Signal Processing (BSP) deals with human behavioral signals. It is defined as processing of human action and behavior data for meaningful analysis to ensure timely decision making and intervention (action) by collaborative integration of human expertise with automated processing. The goal is to support and not supplant humans \cite{NarayananG2013}. The core elements include quantitative understanding of human behavior and mathematical modeling of interaction dynamics. Narayan and Georgiou describe the elements of BSP by using speech and spoken language communication for measuring and modeling human behavior \cite{NarayananG2013}. 

\begin{figure}[htbp]
\includegraphics[width=0.475\textwidth]{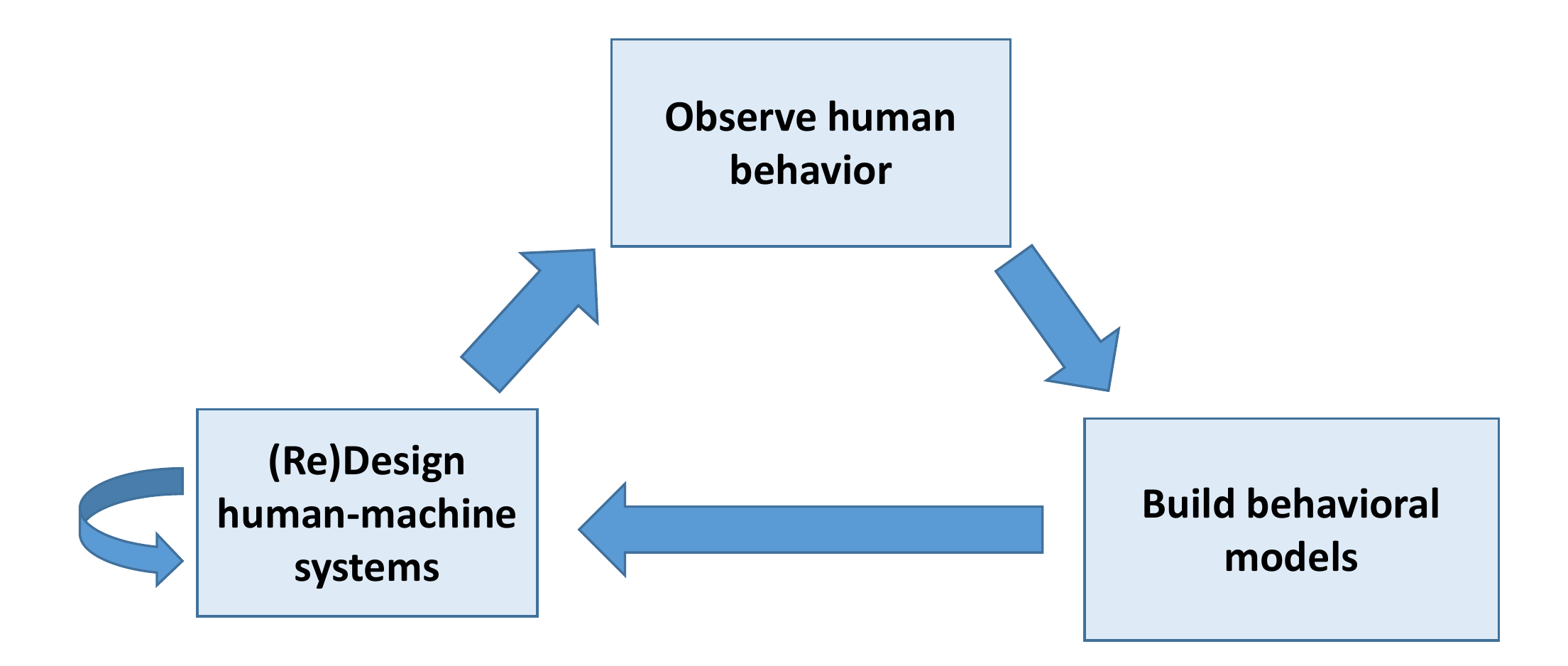}
\caption{General approach for the design and analysis of HuMaINs.}
\label{fig:approach}
\end{figure}

There are two specific research directions while developing BSP algorithms for HuMaINs: 
\begin{enumerate}
\item Develop mathematical models of human decision making using statistical modeling techniques, in close collaboration with cognitive psychologists, and 
\item Design robust fusion algorithms that handle unreliable data from the agents as modeled by the above developed models.
\end{enumerate}
These problems have both theoretical and implementation challenges. Both these research problems are further discussed in some detail below.

\subsection{Statistical Modeling of Human Behavior}
\label{sec:stats}
The first step towards developing efficient systems containing humans and machines is to develop appropriate models that characterize their behavior. While statistical models exist that characterize the machine observations, researchers have not extensively investigated the modeling of decisions and subjective confidences on multi-hypothesis tasks, or on tasks in which human decision makers can provide imprecise (i.e., vague) decisions. Both of these task types, however, are important in the many applications of HuMaINs. In the preliminary work \cite{VempatyVKCV2015}, a comparative study between people and machines for the task of decision fusion has been performed. It was observed that the behavior between people and machines is different since the optimal fusion rule is a deterministic one while people typically use non-deterministic rules which depend on various factors. Based on these observations, a hierarchical Bayesian model was developed to address the observed behavior of humans. This model captured the differences observed in people at individual level, crowd level, and population level. Moving forward, for individual human decision-making models, tools from bounded rationality framework \cite{Simon1982} and rational inattention theory \cite{Sims2003} can be used in building a theory. Experiments with human subjects can be designed to model the cognitive mechanisms which govern the generation of decisions and decision confidences as they pertain to the formulation of precise and imprecise decisions. One can also build models that consider the effect of stress, anxiety, and fatigue in the cognitive mechanisms of human decision making, decision confidence assessment, and response time (similar to \cite{WhiteRVM2010,RatcliffV2011}).

\subsection{Design of Robust Algorithms}

The next step after deriving probabilistic models of human decision-making, is to develop efficient fusion algorithms for collaborative decision making. The goal would be to seek optimal or near-optimal fusion rules which incorporate the informational nature of both humans and machines. Due to the large volume of data in some practical applications, it is also of interest to analyze the effects that a large number of agents (humans/machines) and a high rate of incoming data have on the performance of the fusion rules.  However, the highly parameterized nature of these human models might deem their implementation impractical. Also, the presence of unreliable components in the system might result in poor fusion performance. Data from existing studies in the cognitive psychology literature along with models resulting from the work in Sec.~\ref{sec:stats} can be used in the analysis of these operators. For cases in which the implementation of the optimal rule is not feasible, one must investigate the use of adaptive fusion rules that attempt to learn the parameters of the optimal fusion rule online. Also, for the design of simple and robust algorithms, ideas from coding theory can be used similar to the reliable crowdsourcing results such as in \cite{VempatyVV2014}.

For the development of future systems consisting of humans and machines, the methodology described above needs to be implemented. First, statistical models of humans should be developed, which are then used to optimize the machines in the system. Due to the presence of potential unreliable agents, one has to also take into consideration the robustness of the systems while developing such large-scale systems. For example, \cite{VempatyHV2014,VempatyVV2014,KailkhuraVV2015} demonstrated the utility of statistical learning techniques and tools from coding theory to achieve reliable performance from unreliable agents.

\section{Applications}
Another major driver for the development of large-scale HuMaINs is the application areas. Each application area that deals with human-machine collaboration has its own specific nuances that drive the architecture and algorithmic solutions. In this paper, we discuss four extremely important and timely application areas: education, autonomous vehicles, health-care, and science. We discuss their associated research problems in the context of HuMaINs.

\subsection{Education}
Human-in-the-loop system can have a significant impact in education domain. The research field of Intelligent Tutoring Systems (ITS) is attempting to design computer systems that can provide immediate and customized instruction or feedback to learners, with intervention from a human teacher. They are enabled to serve as complementing a human teacher and ensure personalized and adaptive learning at scale to every learner. While ITS research has been active for several decades \cite{AndersonBR1985,CorbettKA1997}, recent advancements in AI and big data research has enabled increasingly more human-like interactions with computers giving rise to interactive, engaging, and immersive tutoring systems. A typical ITS consists of four basic components \cite{Nwana1990,NkambouMB2010}: Domain model, Student model, Tutoring model, and User interface model. The Domain Model contains the skills, concepts, rules, and/or problem-solving strategies of the domain to be learned. The Student/Learner Model is an overlay on the domain model and it models the student's cognitive and affective understanding of the domain and their evolution during the learning process. The tutor model represents the tutoring strategies and actions that are dependent on the domain model and the specific learner. The user interface component integrates the other three to ensure interaction with the user and learning advances as planned.

With respect to the HuMaIN paradigm discussed in Sec.~\ref{sec:humain}, the domain model represents the task or goal of a HuMaIN, research on the user/learner model represents the human aspect of HuMaINs  and modeling human behavior, the tutoring model represents the machine aspect of HuMaINs and designing of robust inference algorithms, and the user interface model represents the architectural research of designing HuMaINs.

\subsection{Autonomous Vehicles}
Detection, localization, control, and path planning are essential components of autonomous vehicle design~\cite{LevinsonEtAl2011,VivacquaVM2017}. These tasks focus on sensing and interacting with the physical world through sensors and actuators. Although, autonomous vehicles can be a game changer, there are still many obstacles holding back their deployment in practice. Autonomous nature of these systems make them quite vulnerable to cyber-attacks. A solution is to employ human-in-the-loop systems (semi-autonomous driving) for safe and intelligent autonomous vehicle operation. Such systems would require joint environment-driver state sensing, inference, and shared control and new metrics to characterize safety. 
The measures of system safety should take into account human performance in response to unexpected hazardous events, and human decision making during vehicle malfunctions caused by cyber-attacks. Furthermore, allowing communication among multiple self-driving cars can enable collective intelligence in such systems, however, would require the design of robust communication protocols.

\subsection{Health Informatics}
Automated inference using machine learning (ML) for healthcare holds enormous potential to increase quality, efficacy and efficiency of treatment and care~\cite{RaghupathiR2014}. Automatic approaches greatly benefit from big data with many training samples. Several tasks in medical domain have high
dimensional complex data, where the inclusion of a human is impossible and ML shows impressive results. On the other hand, for certain tasks one is confronted with a small number of data sets or rare events, where ML-approaches suffer from insufficient training samples. 
Furthermore, in healthcare, decisions made by machines can have serious
consequences and necessitate the incorporation of human experts' domain knowledge. There is also a growing trend of litigation requiring the need to bring human in the loop. This makes doctor-in-the-loop systems to be a perfect candidate for healthcare. Designing such systems would require devising ML approaches that can interact with human agents (doctors) and can optimize their learning behavior through these interactions. Furthermore, unlike current black-box like ML approaches, we need interpretable ML models for healthcare so that these systems can become transparent to earn experts' trust and be adopted in their workflow.

\subsection{Scientific Discovery}
Scientific research spans problems and challenges ranging from screening of novel materials with desired performance in material science, optimizing the analysis of the Higgs boson in high energy physics, tracking of extreme weather phenomena in climate science. 
Currently, the role of machines in accelerating science has been limited
to solving a well-defined task where the data and techniques are given to them by the scientists. 
This limits our ability to tackle problems where not only the complexity of the data but the questions and the tasks itself challenge our
human capabilities to make discoveries~\cite{tai}. HuMaINs can play an important role in scientific research, and become crucial as more interdisciplinary science questions are tackled.
This would require advancing machine learning techniques to do independent inquiry, proactive learning, and deliberative
reasoning in the presence of hypotheses, domain knowledge, and insights provided by the scientists.

\section{Conclusion}
In this paper, we presented an overview of human machine inference networks. Specific attention was paid to three main issues: 1) architecture design, 2) inference algorithms, and 3) application areas. A holistic research initiative across these issues is needed to empower this new field of HuMaIN research. Also, moving forward, the social aspect of HuMaINs, with multiple human and machine components interacting, such as in IoT systems is a direction for future research.

\bibliographystyle{IEEEtran}
\bibliography{abrv,conf_abrv,vempaty_lib}
\end{document}